\documentstyle[11pt,seceq,amssymb]{article}

\newcommand{\be}{\begin{equation}}
\newcommand{\ee}{\end{equation}}
\newcommand{\lb}{\label}
\newcommand{\en}{\varepsilon}
\newcommand{\om}{\omega}
\newcommand{\bu}{{\bf u}}
\newcommand{\bx}{{\bf x}}

\newcommand{\tor}{{\Bbb T}^3}

\newcommand{\bPi}{{\mbox{\boldmath $\Pi$}}}
\newcommand{\grad}{{\mbox{\boldmath $\nabla$}}}
\newcommand{\bdot}{{\mbox{\boldmath $\cdot$}}}
\newcommand{\bwun}{{\mbox{\boldmath $1$}}}
\newcommand{\btimes}{{\mbox{\boldmath $\times$}}}

\newcommand{\hl}{\hat{\ell}}
\newcommand{\hbl}{{\mbox{\boldmath $\hat{\ell}$}}}
\newcommand{\bl}{{\mbox{\boldmath $\ell$}}}
\newcommand{\bn}{\hat{{\mbox{\boldmath $\ell$}}}}
\newcommand{\hbt}{\hat{{\mbox{\boldmath $t$}}}}
\newtheorem{Th}{Theorem}

\newtheorem{Cor}{Corollary}

\begin{document}
\title{Local 4/5-Law and Energy Dissipation Anomaly in Turbulence}
\author{Gregory L. Eyink\\{\em Department of Mathematics, University of
Arizona, Tucson, AZ 85721}}
\date{ }
\maketitle
\begin{abstract}
A strong local form of the ``4/3-law'' in turbulent flow has been proved
recently
by Duchon and Robert for a triple moment of velocity increments averaged over
both a bounded spacetime region and separation vector directions, and for
energy
dissipation averaged over the same spacetime region. Under precisely stated
hypotheses,
the two are proved to be proportional, by a constant 4/3, and to appear as a
nonnegative defect measure in the local energy balance of singular
(distributional)
solutions of the incompressible Euler equations. Here we prove that the energy
defect
measure can be represented also by a triple moment of purely longitudinal
velocity
increments and by a mixed moment with one longitudinal and two tranverse
velocity
increments. Thus, we prove that the traditional 4/5- and 4/15-laws of
Kolmogorov hold
in the same local sense as demonstrated for the 4/3-law by Duchon-Robert.
\end{abstract}

\newpage

\section{Introduction}

Recently, Duchon and Robert \cite{DR00} have established an energy balance
relation for distributional solutions of the three-dimensional (3D)
incompressible
Euler equations. Their balance relation contains a ``defect'' or ``anomaly''
term,
with an interesting connection to turbulence theory. Since the work of
Duchon-Robert
provides the point of departure of the present paper, it is appropriate to
describe
their theorems briefly here. For our purposes, there are three main results:

First, if $\bu\in L^3([0,T]\times\tor)$ is a weak solution of the
incompressible Euler
equations on the 3-torus $\tor$, then it is proved in \cite{DR00} (Proposition
2) that
the following local balance holds in the sense of distributions:
\be \partial_t({{1}\over{2}}|\bu|^2)+\grad\bdot [({{1}\over{2}}|\bu|^2
+p)\bu ] = -D(\bu). \lb{Ebal} \ee
Here $D(\bu)$ is a defect distribution which for classical solutions
vanishes identically, implying local energy conservation. Duchon and Robert
also establish various expressions for the defect term. In particular, they
have shown that
\be D(\bu) = {\mathcal D}{\rm -}\lim_{\!\!\!\!\!\!\!\!\!\!\!\!\!\!\!\!
                            \en\rightarrow 0}{{1}\over{4}}\int_{\tor}
\grad\varphi_\en(\bl)\bdot\delta\bu(\bl)|\delta\bu(\bl)|^2\,\,d^3\bl
    \lb{defect} \ee
where ${\mathcal D}{\rm -}\lim$ means limit in the sense of distributions on
$[0,T]\times \tor$, with $\varphi\in C^\infty_0(\tor),$ even, nonnegative
with unit integral, $\varphi^\en(\bx)= \en^{-3}\varphi(\bx/\en)$, and
$\delta\bu(\bx,t;\bl)= \bu(\bx+\bl,t)-\bu(\bx,t)$. This expression is
remarkable
because it is closely connected with an exact result in turbulence theory,
the so-called ``K\'{a}rm\'{a}n-Howarth-Monin relation'':
\be \grad\bdot\langle \delta\bu(\bl)|\delta\bu(\bl)|^2\rangle = -4\bar{\en}
    \lb{Yaglom}. \ee
See \cite{Frisch}, section 6.2.1. In this relation, $\bar{\en}=
\nu\langle|\grad\bu^\nu|^2\rangle$
is the mean energy dissipation for a Navier-Stokes solution $\bu^\nu$, which is
assumed to remain finite
as viscosity $\nu$ tends to zero. We see that this result is essentially
equivalent to the statement that
the Duchon-Robert defect satisfy $\langle D\rangle =\bar{\en}$.

For a general distributional solution of Euler, there need be no connection of
the defect with
viscous dissipation, nor need it even be true that $D(\bu)\geq 0$. However, a
second theorem (Proposition 4)
in \cite{DR00} states that, if $\bu^\nu$ is a Leray solution of incompressible
Navier-Stokes
equation for viscosity $\nu$ and if $\bu^\nu\rightarrow\bu$ strong in
$L^3([0,T]\times\tor)$
as $\nu\rightarrow 0$, then
\be D(\bu) =  {\mathcal D}{\rm -}\lim_{\!\!\!\!\!\!\!\!\!\!\!\!\!\!\!\!
\nu\rightarrow 0} [ \nu|\grad\bu^\nu|^2+ D(\bu^\nu)] \lb{nu-defect} \ee
Since it is well-known that $D(\bu^\nu)\geq 0$ for Leray solutions, thus also
$D(\bu)$ is a nonnegative distribution, i.e. a Radon measure (\cite{Choq},
Example 12.5).
This theorem may be paraphrased as saying that strong viscosity solutions of
the incompressible Euler
equations are also dissipative solutions.

Finally, Duchon and Robert, under an additional hypothesis, establish an even
simpler form of
the defect distribution. With $\omega$ unit Haar measure on $S^2$, they define
the function
\be  S(\bu,\ell)(\bx,t):= {{1}\over{\ell}}\int_{S^2} d\om(\bn) \delta
u_L(\bx,t;\bl)|\delta\bu(\bx,t;\bl)|^2. \lb{Su} \ee
in $L^1([0,T]\times\tor)$. Here $\delta
u_L(\bx,t;\bl)=\bn\bdot\delta\bu(\bx,t;\bl)$ is the longitudinal velocity
increment.
Assuming that the following limit exists
\be S(\bu)(\bx,t) := {\mathcal D}{\rm -}\lim_{\!\!\!\!\!\!\!\!\!\!\!\!\!\!\!\!
\ell\rightarrow 0} S(\bu,\ell)(\bx,t) \lb{su} \ee
Duchon and Robert \cite{DR00}, Section 5 show that
\be    S(\bu) = -{{4}\over{3}}D(\bu).    \lb{Du} \ee
This is a rigorous form of another well-known relation in turbulence theory,
sometimes called the ``4/3-law'':
\be \langle \delta u_L(\bl) |\delta\bu(\bl)|^2\rangle \sim
-{{4}\over{3}}\bar{\en}\ell. \lb{4-3law} \ee
See \cite{AOAZ}. It is well-known that the K\'{a}rm\'{a}n-Howarth-Monin
relation reduces to the 4/3-law
under conditions of local isotropy. This is achieved here by the angle average
over the sphere in (\ref{Su}).

Such relations as (\ref{Yaglom}) and (\ref{4-3law}) in turbulence theory go
back to the original work of
A. N. Kolmogorov \cite{K41}. However, Kolmogorov in fact proved a relation
involving only longitudinal
velocity increments, the so-called ``4/5-law'':
\be \langle [\delta u_L(\bl)]^3\rangle \sim -{{4}\over{5}}\bar{\en}\ell
\lb{4-5law} \ee
This was established from the Navier-Stokes equations, under conditions of
statistical homogeneity and local
isotropy and with the assumption that energy dissipation remains finite in the
zero-viscosity limit.
It is our purpose here to establish an expression for the Duchon-Robert energy
dissipation anomaly $D(\bu)$
with exactly the form of Kolmogorov's law. Our proof yields another expression
which is related to the
so-called ``4/15-law'':
\be \langle \delta u_L(\bl) |\delta u_T(\bl)|^2\rangle \sim
-{{4}\over{15}}\bar{\en}\ell \lb{4-15law} \ee
Here $\delta u_T(\bl)=\hbt\bdot\delta\bu(\bl)$ is a transverse velocity
increment, with $\hbt$
any unit vector orthogonal to $\bn$. This relation is known to hold under the
same conditions as the 4/5-law.
The rigorous derivation of the 4/5- and 4/15-laws given here, under precisely
stated assumptions, yields a result with
a wider domain of validity than that of some previous rigorous derivations,
such as that of Nie and Tanveer \cite{NieTan}.
In particular, the form of the 4/5-law established here---like the
Duchon-Robert version of the 4/3-law---is
{\it local}, in the sense that it relates third-order moments of velocity
increments and viscous dissipation
averaged over the same bounded spacetime region, not necessarily large.

In the following section we prove our main theorem. In a final discussion
section we discuss briefly
its physical significance and a compare it with previous results. Let us now
make just a few remarks
on notations: The symbol $L^p$ will be used below for $L^p([0,T]\times\tor)$.
If $F(\bx,t)$
is any spacetime distribution, we denote $F^\en(\bx,t)=(\varphi^\en*F)(\bx,t)$,
where
$\varphi\in C_0^\infty(\tor)$ and $*$ is space convolution. Thus, $F^\en$
remains, for each fixed $\bx$,
a distribution in $t$. Often below, as above, we omit the variables $(\bx,t)$
where their presence
is clear from the context.

\newpage

\section{The Main Theorem}

We prove the following:

\begin{Th}
Let $\bu\in L^3([0,T]\times\tor)$ be a weak solution of the incompressible
Euler equations
on the 3-torus $\tor$. Let $\varphi(\ell)$ be any $C^\infty$ function with
compact support,
nonnegative with unit integral, spherically symmetric, and let
$\varphi^\en(\ell)
=\en^{-3}\varphi(\ell/\en)$. Finally, define longitudinal and transverse
velocity
increments as
\be \delta \bu_L(\bx,t;\bl) = (\hbl\otimes\hbl)\delta\bu(\bx,t;\bl), \,\,\,\,
    \delta \bu_T(\bx,t;\bl) = (\bwun-\hbl\otimes\hbl)\delta\bu(\bx,t;\bl),
\lb{delta-u-LT} \ee
Then, the following functions in $L^1([0,T]\times\tor)$
\be D_L^\en(\bu) = {{3}\over{4}}\int_{\tor} d^3\bl \,\,\left\{
       \grad\varphi^\en(\ell)\bdot\delta\bu(\bl)|\delta \bu_L(\bl)|^2 +
       {{2}\over{\ell}}\varphi^\en(\ell)\delta
u_L(\bl)|\delta\bu_T(\bl)|^2\right\}, \lb{DL} \ee
and
\be D_T^\en(\bu) = {{3}\over{8}}\int_{\tor} d^3\bl \,\,\left\{
       \grad\varphi^\en(\ell)\bdot\delta\bu(\bl)|\delta \bu_T(\bl)|^2 -
       {{2}\over{\ell}}\varphi^\en(\ell)\delta
u_L(\bl)|\delta\bu_T(\bl)|^2\right\}, \lb{DT} \ee
both converge in the sense of distributions as $\en\rightarrow 0$ to $D(\bu)$,
where the latter
is the defect distribution in the local energy balance for $\bu$,
\be \partial_t({{1}\over{2}}|\bu|^2)+\grad\bdot [({{1}\over{2}}|\bu|^2 +p)\bu ]
= -D(\bu),
\lb{Ebal-Th} \ee
established earlier by Duchon-Robert.
\end{Th}

We shall prove this theorem in several steps.

The idea of the proof is to consider separate balance equations for the
longitudinal and
transverse components of the energy. We define first longitudinal and
transverse velocities
relative to a vector $\bl$:
\be \bu_L(\bx,t;\bl) = (\hbl\otimes\hbl)\bdot \bu(\bx+\bl,t),\,\,\,\,
    \bu_T(\bx,t;\bl) = (\bwun-\hbl\otimes\hbl)\bdot \bu(\bx+\bl,t). \lb{u-LT}
\ee
Of course, $\bu_L+\bu_T=\bu$. We define also mollified versions
\be \bu^\en_X(\bx,t) = \int_{\tor} d^3\bl \,\,\varphi^\en(\ell)
\bu_X(\bx,t;\bl), \,\,\,\, X=L,T
     \lb{moll-u-LT}\ee
It is easy to see that these satisfy the equations
\be \partial_t \bu^\en_X +\grad\bdot (\bu\otimes\bu_X)^\en =
-\grad\bdot\bPi^\en_X, \,\,\,\, X=L,T
     \lb{eq-u-LT} \ee
distributionally in time, where
\be \bPi^\en_L(\bx,t)
     = \int_{\tor} d^3\bl \,\,\varphi^\en(\ell) (\hbl\otimes\hbl)
p(\bx+\bl,t),\,\,\,\,
    \bPi^\en_T(\bx,t)
     = \int_{\tor} d^3\bl \,\,\varphi^\en(\ell) (\bwun-\hbl\otimes\hbl)
p(\bx+\bl,t) \lb{Pi-LT}. \ee

These equations can be simplified by the observation that
\be \grad\bdot\bPi^\en_X(\bx,t) = \grad p_X^\en(\bx,t), \,\,\,\, X=L,T
\lb{Pi-p-LT} \ee
where $p_L^\en,p_T^\en$ are scalar functions defined by
\be p_X^\en(\bx,t) = \int_{\tor} d^3\bl \,\,\varphi^\en_X(\ell) p(\bx+\bl,t),
\,\,\,\, X=L,T
    \lb{p-LT} \ee
with
\be \varphi_L(\ell) = \varphi(\ell)-\varphi_T(\ell), \,\,\,\,
    \varphi_T(\ell) = 2 \int_\ell^\infty {{\varphi(\ell')}\over{\ell'}}
\,\,d\ell'.
    \lb{phi-LT} \ee
Note that $\varphi_L,\varphi_T$ are compactly supported and $C^\infty$
everywhere except at 0,
where they have a mild (logarithmic) singularity. To prove (\ref{Pi-p-LT}), we
use the elementary
relation $\grad \hbl= \bwun-\hbl\otimes\hbl$. A simple computation then gives,
for example,
\be \grad \bdot \bPi^\en_L(\bx,t)
     = -\int_{\tor} d^3\bl \,\,\left\{ {{d\varphi^\en}\over{d\ell}}(\ell)
        + {{2}\over{\ell}}\varphi^\en(\ell)\right\} \hbl
\,\,p(\bx+\bl,t),\,\,\,\, \lb{grad-Pi-L} \ee
{}From its definition,
\be \grad \varphi_L(\ell) = \left\{ {{d\varphi^\en}\over{d\ell}}(\ell) +
{{2}\over{\ell}}
                                            \varphi^\en(\ell)\right\} \hbl.
\lb{grad-phi-L} \ee
This gives easily $\grad\bdot\bPi^\en_L = \grad p_L^\en$ in the sense of
distributions. Because
$\bPi^\en_L+\bPi^\en_T = p^\en \bwun$ and $p_L^\en + p_T^\en=p^\en$, this
yields also the
relation $\grad\bdot\bPi^\en_T = \grad p_T^\en$. Finally, we obtain the simpler
equations
\be \partial_t \bu^\en_X +\grad\bdot (\bu\otimes\bu_X)^\en = -\grad p^\en_X,
\,\,\,\, X=L,T
     \lb{simp-eq-u-LT} \ee
for $\bu^\en_L$ and $\bu^\en_T$.

We next observe that both $\bu^\en_L$ and $\bu^\en_T$ are divergence-free. In
fact, a computation
like that above shows that
\begin{eqnarray}
\grad \bdot \bu^\en_L(\bx,t)
    & = & -\int_{\tor} d^3\bl \,\,\left\{ {{d\varphi^\en}\over{d\ell}}(\ell)
          + {{2}\over{\ell}}\varphi^\en(\ell)\right\} \hbl\bdot \bu(\bx+\bl,t),
\cr
  \,& = & -\int_{\tor} d^3\bl \,\, \grad\varphi^\en_L(\ell) \bdot
\bu(\bx+\bl,t). \lb{grad-u-L}
\end{eqnarray}
If we integrate against any smooth test function $\psi(\bx)$, we then get
\be \int_{\tor} d^3\bx\,\,\psi(\bx) \grad \bdot \bu^\en_L(\bx,t)
    = \int_{\tor} d^3\bl \,\, \varphi^\en_L(\ell) \grad\bdot (\psi*\bu)(\bl,t).
\lb{div-u-L} \ee
The latter is zero, since $\bu$ is divergence-free (in the distributional
sense). Thus,
$\grad \bdot \bu^\en_L=0$. Since $\bu^\en_L +\bu^\en_T=\bu^\en,$ then also
$\grad \bdot \bu^\en_T=0$.

{}From the equations (\ref{simp-eq-u-LT}) for $X=L,T$, the incompressibility
conditions
for $\bu^\en_L$ and $\bu^\en_T$, and the Euler equations for $\bu$ (in
distribution sense),
we derive the following balance equations
\be 2\partial_t (\bu\bdot\bu^\en_L) +\grad\bdot[ 2(\bu\bdot\bu^\en_L)\bu +
   ((\bu_L\bdot\bu_L)\bu)^\en-(\bu_L\bdot\bu_L)^\en\bu + 2p\bu_L^\en +
2p_L^\en\bu]
   = -{{4}\over{3}}D_L^\en(\bu), \lb{uuL-eq} \ee
and
\be 2\partial_t (\bu\bdot\bu^\en_T) +\grad\bdot[ 2(\bu\bdot\bu^\en_T)\bu +
   ((\bu_T\bdot\bu_T)\bu)^\en-(\bu_T\bdot\bu_T)^\en\bu + 2p\bu_T^\en +
2p_T^\en\bu]
   = -{{8}\over{3}}D_T^\en(\bu). \lb{uuT-eq} \ee
The basic identities used to derive these equations are
\begin{eqnarray}
\, & & \int_{\tor} d^3\bl \,\,\left\{
       \grad\varphi^\en(\ell)\bdot\delta\bu(\bl)|\delta \bu_L(\bl)|^2 +
       {{2}\over{\ell}}\varphi^\en(\ell)\delta
u_L(\bl)|\delta\bu_T(\bl)|^2\right\} \cr
\, & & \,\,\,\,\,\,\,\,\,\,\,\,\,\,\,\,=\int_{\tor} d^3\bl
\,\,{{\partial}\over{\partial\ell_k}}
       \{\hl_i\hl_j\varphi^\en(\ell)\} \delta u_i(\bl)\delta u_j(\bl)\delta
u_k(\bl) \cr
\, & & \,\,\,\,\,\,\,\,\,\,\,\,\,\,\,\,= -{{\partial}\over{\partial x_k}}
       [((\bu_L\bdot\bu_L)u_k)^\en-(\bu_L\bdot\bu_L)^\en u_k]
       +2 u_i{{\partial}\over{\partial x_k}}[((u_{Li} u_k)^\en- u_{Li}^\en
u_k], \lb{id-L}
\end{eqnarray}
and
\begin{eqnarray}
\, & & \int_{\tor} d^3\bl \,\,\left\{
       \grad\varphi^\en(\ell)\bdot\delta\bu(\bl)|\delta \bu_T(\bl)|^2 -
       {{2}\over{\ell}}\varphi^\en(\ell)\delta
u_L(\bl)|\delta\bu_T(\bl)|^2\right\} \cr
\, & & \,\,\,\,\,\,\,\,\,\,\,\,\,\,\,\,=\int_{\tor} d^3\bl
\,\,{{\partial}\over{\partial\ell_k}}
       \{(\delta_{ij}-\hl_i\hl_j)\varphi^\en(\ell)\} \delta u_i(\bl)\delta
u_j(\bl)\delta u_k(\bl) \cr
\, & & \,\,\,\,\,\,\,\,\,\,\,\,\,\,\,\,= -{{\partial}\over{\partial x_k}}
       [((\bu_T\bdot\bu_T)u_k)^\en-(\bu_T\bdot\bu_T)^\en u_k]
       +2 u_i{{\partial}\over{\partial x_k}}[((u_{Ti} u_k)^\en-u_{Ti}^\en u_k].
\lb{id-T}
\end{eqnarray}

Now we study the convergence of the lefthand sides of (\ref{uuL-eq}) and
(\ref{uuT-eq})
as $\en\rightarrow 0$.

First we show that $\bu_L^\en\rightarrow {{1}\over{3}}\bu$  and
$\bu_T^\en\rightarrow {{2}\over{3}}\bu$
strong in $L^3$ as $\en\rightarrow 0$. Because of the spherical symmetry,
compact support and
unit integral of $\varphi^\en$,
\be \int_{\tor} d^3\bl \,\,\varphi^\en(\ell)\hbl\otimes\hbl  =
{{1}\over{3}}\bwun, \lb{third-delta} \ee
for sufficiently small $\en$. From this result and the definition of
$\bu_L^\en$ it follows that
\be \bu_L^\en(\bx,t)-{{1}\over{3}}\bu(\bx,t)=\int_{\tor} d^3\bl
\,\,\varphi^\en(\ell)\hbl\otimes\hbl\bdot
           [\bu(\bx+\bl,t)-\bu(\bx,t)]. \lb{uL-thirdu} \ee
Thus,
\be \|\bu_L^\en-{{1}\over{3}}\bu\|_{L^3} \leq \int_{\tor} d^3\bl
\,\,\varphi^\en(\ell)
           \|\bu(\cdot+\bl)-\bu\|_{L^3}. \lb{norm-uL-thirdu} \ee
Because $\bu\in L^3$, it follows by a standard approximation argument that
$\|\bu(\cdot+\bl)-\bu\|_{L^3}
\rightarrow 0$ as $\ell\rightarrow 0$. Hence, it follows from
(\ref{norm-uL-thirdu}) that
$\|\bu_L^\en-{{1}\over{3}}\bu\|_{L^3}\rightarrow 0$ as $\en\rightarrow 0$, as
was claimed. Because
$\bu_T^\en=\bu^\en-\bu_L^\en$, we also deduce that
$\|\bu_T^\en-{{2}\over{3}}\bu\|_{L^3}\rightarrow 0$.

Next we show that $p_L^\en\rightarrow {{1}\over{3}}p$  and $p_T^\en\rightarrow
{{2}\over{3}}p$
strong in $L^{3/2}$ as $\en\rightarrow 0$. First, observe that
$p=(-\bigtriangleup)^{-1}
\partial_i\partial_j(u_iu_j)$, so that $p\in L^{3/2}$ by the
Calder\'{o}n-Zygmund inequality.
{}From the definitions of $\phi_T$ and $p_T^\en$ it follows easily that
\be p_T^\en(\bx,t) = {{2}\over{3}}\cdot 4\pi\int_0^\infty \ell'^2
d\ell'\varphi^\en(\ell')
                   \cdot {{1}\over{|B_{\ell'}|}}\int_{B_{\ell'}} d^3\bl
\,\,p(\bx+\bl,t), \lb{pT-eq} \ee
where $B_\ell$ is the ball of radius $\ell$ centered at the origin and
$|B_\ell|$ is its volume.
Because $\varphi^\en$ has unit integral,
\be p_T^\en(\bx,t)-{{2}\over{3}}p(\bx,t) = {{2}\over{3}}\cdot 4\pi\int_0^\infty
\ell'^2 d\ell'
    \varphi^\en(\ell')
    \cdot {{1}\over{|B_{\ell'}|}}\int_{B_{\ell'}} d^3\bl
\,\,[p(\bx+\bl,t)-p(\bx,t)], \lb{pT-twothirdp} \ee
and thus
\be \|p_T^\en-{{2}\over{3}}p\|_{L^{3/2}} \leq {{2}\over{3}}\cdot
4\pi\int_0^\infty \ell'^2 d\ell'
    \varphi^\en(\ell') \cdot {{1}\over{|B_{\ell'}|}}\int_{B_{\ell'}} d^3\bl
\,\,\|p(\cdot+\bl)-p\|_{L^{3/2}}.
    \lb{norm-pT-twothirdp} \ee
Consequently, $\|p_T^\en-{{2}\over{3}}p\|_{L^{3/2}}\rightarrow 0$ as
$\en\rightarrow 0$, as was
to be proved. Since $p_L^\en=p^\en-p_T^\en$, also
$\|p_L^\en-{{1}\over{3}}p\|_{L^{3/2}}\rightarrow 0$.

Entirely similar arguments show that
$((\bu_X\bdot\bu_X)\bu)^\en-(\bu_X\bdot\bu_X)^\en\bu\rightarrow 0$
strong in $L^1$ as $\en\rightarrow 0$, for $X=L,T.$ We leave that to the
reader.

Finally we see that the the lefthand side of equation (\ref{uuL-eq}) converges
as $\en\rightarrow 0$ to
the quantity
$$ {{4}\over{3}}\left\{\partial_t({{1}\over{2}}|\bu|^2)
+\grad\bdot\left[({{1}\over{2}}|\bu|^2+p)\bu\right]
\right\} $$
in the sense of distributions, and that the lefthand side of (\ref{uuT-eq})
converges
to ${{8}\over{3}}$ times the same quantity in the curly brackets, also in the
distribution sense.
However, by the result of Duchon-Robert \cite{DR00}, that quantity in the
brackets is equal to minus
the defect distribution $D(\bu)$. We conclude then that
\be D_X^\en(\bu)\rightarrow D(\bu) \lb{L-conv} \ee
as $\en\rightarrow 0$ for both $X=L,T$, in the sense of distributions. $\Box$

We now prove the following:
\begin{Cor}
Assume that the functions $S_L(\bu,\ell),S_T(\bu,\ell)\in L^1([0,T]\times\tor)$
defined by
\be S_L(\bu,\ell)= {{1}\over{\ell}}\int_{S^2} d\om(\hbl) [\delta u_L(\bl)]^3
\lb{Su-L} \ee
and
\be S_T(\bu,\ell)= {{1}\over{\ell}}\int_{S^2} d\om(\hbl) \delta u_L(\bl)|\delta
\bu_T(\ell)|^2 \lb{Su-T} \ee
have limits
\be S_X(\bu) = {\mathcal D}{\rm -}\lim_{\!\!\!\!\!\!\!\!\!\!\!\!\!\!\!\!
               \ell\rightarrow 0} S_X(\bu,\ell), \,\,\,\, X=L,T. \lb{lim-Su-LT}
\ee
Then,
\be S_L(\bu) = -{{4}\over{5}} D(\bu),\,\,\,\, S_T(\bu)=-{{8}\over{15}} D(\bu).
\lb{laws} \ee
\end{Cor}

The proof is quite straightforward. An easy computation for a spherically
symmetric test function gives
\be {{4}\over{3}} D_L^\en(\bu) = 4\pi \int_0^\infty d\ell\,\,\left[
\ell^3\varphi'(\ell) S_L(\bu,\en\ell)
                               + 2 \ell^2 \varphi(\ell) S_T(\bu,\en\ell)\right]
\lb{en-L-eq} \ee
and
\be {{8}\over{3}} D_T^\en(\bu) = 4\pi \int_0^\infty d\ell\,\,\left[
\ell^3\varphi'(\ell)
                               - 2\ell^2 \varphi(\ell) \right]
S_T(\bu,\en\ell). \lb{en-T-eq} \ee
Taking the limit as $\en\rightarrow 0$, using Theorem 1, the normalization of
the test function,
and the above hypotheses, we see that
\be {{4}\over{3}} D(\bu) =  -3 S_L(\bu) + 2 S_T(\bu) \lb{L-eq} \ee
and
\be {{8}\over{3}} D(\bu) = -5 S_T(\bu). \lb{T-eq} \ee
Solving this linear system gives the result. $\Box$

In the definition of $S_T(\bu,\ell)$ in (\ref{Su-T}) we could instead have
replaced $|\delta\bu_T|^2$
by the square magnitude of a transverse component $\delta u_T=\hbt\bdot
\delta\bu$, where
$\hbt$ is any unit vector perpendicular to $\bn$. Call such a quantity
$\tilde{S}_T(\bu,\ell)$.
However, $|\delta\bu_T|^2=|\delta u_T|^2+|\delta u_T'|^2,$ where $\delta
u_T'=(\hbt\btimes\bn)\bdot\delta\bu.$
Since both transverse components give an equal contribution in the spherical
average, $S_T(\bu,\ell)=2
\tilde{S}_T(\bu,\ell).$ Under the assumptions of the corollary,
$\tilde{S}_T(\bu,\ell)$ then also has a limit
distribution $\tilde{S}_T(\bu)$ as $\ell\rightarrow 0$, and
\be  \tilde{S}_T(\bu)=-{{4}\over{15}} D(\bu). \lb{conv-4-5law} \ee
This is the conventional statement of the ``4/15-law''.

\section{Discussion}

If we couple our Corollary 1 with Duchon-Robert's characterization of $D(\bu)$
by the inviscid limit,
\cite{DR00}, Proposition 4, then we arrive at essentially the following
statement: For any ``nice'' spacetime
region $R\subset\tor \times [0,T]$, let
\be    \en_R = \lim_{\nu\rightarrow 0}{{1}\over{|R|}}\int\int_R d^3\bx
\,\,dt\,\,\en(\bx,t) \lb{B-av} \ee
with $\en(\bx,t):=\nu|\grad\bu^\nu|^2$ (assuming for simplicity that
$D(\bu^\nu)=0$).
If first $\nu\rightarrow 0$ and next $\ell\rightarrow 0$, then
\be \langle [\delta u_L(\bl)]^3 \rangle_{ang,R}
            \sim -{{4}\over{5}} \en_R\ell \lb{Hyp1-RSH} \ee
with $\langle \cdot\rangle_{ang,R}$ denoting an average of $(\bn,\bx,t)$ over
$S^2\times R$.
To be precise, the relation should hold in the sense of averaging in $\bx,t$
against smooth
test functions $\psi(\bx,t)$ rather than against a characteristic function
$1_R(\bx,t)$. The latter sense is stronger,
because, if the result holds for characteristic functions of Borel sets $R$,
then, by approximation arguments,
it holds also for smooth functions. In fact, by the portmanteau theorem
\cite{Bill}, it is enough to show that it holds
for all bounded Borel sets $R$ such that $D({\partial R})=0$, i.e. with no
energy dissipation concentrating
on the boundary. We see from these remarks that the meaning of the Corollary 1
is that the scaling observed
in the Kolmogorov ``4/5-law'' \cite{K41} should hold with just local averaging
in spacetime and an angular average
over the direction of the separation vector. Note that, if the solution of the
Navier-Stokes equations at finite viscosity
is not regular, then the same relation would hold with the modified definition
$\en(\bx,t):
=\nu|\grad\bu^\nu|^2+D(\bu^\nu)$, including the dissipation from the
singularities.

Another recent work by Nie and Tanveer \cite{NieTan} has given a rigorous
derivation of both the 4/5-
and 4/3-laws under conditions similar to those of Duchon-Robert and of the
present paper. The proof of
those laws by Nie and Tanveer is for individual solutions of the Navier-Stokes
equations (assumed
to be strong) without any statistical averaging, and with no assumptions of
homogeneity or isotropy.
In \cite{NieTan}, as here, the averaging is over spacetime and over directions
of the separation vector.
However, Nie and Tanveer considered the space average over the entire domain
and time average
over an interval $[0,T]$ with $T\rightarrow \infty$. In contrast, the present
result holds for any
spacetime domain, of arbitrary extent, so long as the Reynolds number is
sufficiently high, and also
for solutions of the Navier-Stokes equations that may be singular. In this
respect, our result is stronger
than that of \cite{NieTan}. On the other hand, our proof and that in
\cite{DR00} give no indication
how large the Reynolds number must be taken to approach the limit, whereas Nie
and Tanveer establish
precise error bounds for their result.

There is a slight resemblance of our and Duchon-Robert's local results with the
``refined similarity
hypothesis''that Kolmogorov postulated in 1962 \cite{K62}. Kolmogorov
considered averages $\en_B$
of dissipation over balls $B\subset \tor$ and the structure functions obtained
by averaging
velocity increments $\delta\bu(\bx,t;\bl)$ conditioned upon the local
dissipation averaged over
a ball $B_\ell(\bx)$ of diameter $\ell$ centered at the midpoint
${{1}\over{2}}(\bx+\bl)$.
He postulated that these might be calculated by his original 1941 theory, e.g.
for longitudinal
differences. Hence, he obtained for unconditioned ensemble averages
\be \langle [\delta u_L(\bl)]^{3n}\rangle \sim C_n
                                        \langle\en_{B_\ell}^n\rangle \ell^n
\lb{RSH} \ee
Our scaling relation (\ref{Hyp1-RSH}) is supposed to hold pathwise, for
individual realizations.
Hence, if it holds---along with a uniform and integrable bound for
$\ell\rightarrow 0$---
then one can infer that
\be \langle \left[\langle(\delta u_L(\bl))^3 \rangle_{ang,R}\right]^n
    \rangle \sim \left(-{{4}\over{3}}\right)^n \langle \en_R^n\rangle \ell^n.
\lb{Hyp2-RSH} \ee
The similarity is apparent. Aside from the fact that we must integrate over
time as well as space,
the major and significant difference is that in the relation (\ref{Hyp2-RSH})
the region $R$ must
be fixed as $\ell\rightarrow 0$, while in Kolmogorov's hypothesis the ball
$B_\ell$ itself shrinks
to zero in the limit. Our results and that of Duchon-Robert say nothing about
intermittency.

Although we have proved a rigorous theorem, it is only established under
various hypotheses. We regard
those as plausible, but it is nevertheless of some interest to inquire about
the feasibility of
numerical and/or experimental tests of the local results proved here and in
\cite{DR00}. In \cite{Emp1},
the 4/5-law has been verified both in a numerical simulation via a
space-average over the domain
and also in experimental hot-wire data. However, in the same simulation, the
4/15-law is not satisfied
(S. Chen, private communication). Similar violations of the 4/15-law were
observed in experimental X-wire data
from an atmospheric boundary layer \cite{KurSreen}. In both cases, anisotropy
of the flow may be suspected
of vitiating the result. In that case, angle-averaging as employed in the
rigorous theorems may improve the agreement.
On the other hand, a more recent numerical study \cite{GFN} with a $1024^3$
simulation has found that the regimes
of validity of both the 4/5- and 4/3-laws, in even their global form, are
established only very slowly as
the Reynolds number is increased. Such a slow approach to asymptotia makes a
direct test of the local
results, especially a verification of the numerical prefactor, rather more
difficult.

\newpage

\noindent {\bf Acknowledgements:} This work, LAUR 02-4761, was carried out in
part at the Los Alamos
National Laboratory under the auspices of the Department of Energy and
supported by LDRD-ER 2000047.
The work was begun at the Erwin Schroedinger International Institute for
Mathematical Physics
during its Program on Developed Turbulence in spring 2002. I wish to
acknowledge several of my
colleagues there for fruitful discussions, in particular Jean Duchon, Susan
Kurien, and Christos
Vassilicos. I would also like to thank Mark Taylor of Los Alamos National
Laboratory and Shiyi Chen
at Johns Hopkins University for information about their numerical simulations.

\end{document}